\documentclass[12pt]{article}

\let\LARGE=\Large
\let\Large=\large
\let\large=\normalsize
\newcommand{\g}{\gamma}

\newcommand{\ft}[2]{{\textstyle\frac{#1}{#2}}}
\newcommand{\be}{\begin{equation}}
\newcommand{\eq}{\end{equation}}

\def\beqa{\begin{eqnarray}}
\def\eeqa{\end{eqnarray}}

\begin {document}
\begin{titlepage}
\begin{center}
\hfill SU-ITP-04/38\\
\hfill SLAC-PUB-10790\\
\hfill TIFR/TH/04-27\\
%\hfill {\tt hep-th/0410076}\\

\vskip 2cm

%{ \LARGE \bf    Quantum Gravity Clothing for a Classical Singularity}

%{\LARGE \bf Quantum Horizons for Null Singularities}

%{\LARGE \bf Quantum Horizons for Classical Singularities}

%{\LARGE \bf Quantum Gravity Horizons for Null Singularities}

%{\LARGE \bf Stringy Clothes for Classical Singularities}

{ \LARGE \bf  A Stringy Cloak for a Classical Singularity}

\vskip 1cm

{\bf Atish Dabholkar$^{1}$\footnote{\mbox{
Email: {\tt atish@tifr.res.in} }},
Renata Kallosh$^{2}$\footnote{\mbox{
Email: {\tt kallosh@stanford.edu} }}
and Alexander Maloney$^{2,3}$\footnote{\mbox{
Email: {\tt maloney@slac.stanford.edu}
}}}
\\

\vskip 1cm

%\vspace{.5truecm}

{\em }

\centerline{$^{1}$ \it Department of Theoretical Physics,}
\centerline{\it Tata Institute of Fundamental Research}
\centerline{\it Homi Bhabha Road, Mumbai 400005, India}
\centerline{}

\centerline{$^2$ \it Institute for Theoretical Physics, Department
of Physics} \centerline{\it Stanford University, Stanford, CA
94305} \centerline{}

\centerline{$^3$ \it Theory Group, Stanford Linear Accelerator
Center} \centerline{\it 2575 Sand Hill Rd, Menlo Park, CA 94025}
\end{center}

\vskip 1cm

\begin{center} {\bf ABSTRACT } \end{center}
%%%%%%%%%%%%%%%%%%%%%%%%%%%%%%%%%%%%%%%%%%%%%%%%%%%%%%%

We consider a class of 4D supersymmetric black hole solutions,
arising from string theory compactifications, which classically
have vanishing horizon area and singular space-time geometry.
String theory motivates the inclusion of higher derivative terms,
which  convert these singular classical solutions into regular
black holes with finite horizon area. In particular, the
supersymmetric attractor equations imply that the central charge,
which determines the radius of the $AdS_2\times S^2$ near horizon
geometry, acquires a non-vanishing value due to quantum effects.
In this case quantum corrections to the Bekenstein-Hawking relation
between entropy and area are large. 
This is the first explicit example where stringy quantum gravity
effects replace a classical  null singularity by a black hole with
finite horizon area.

\vfill

October 2004\\
\end{titlepage}

%%%%%%%%%%%%%%%%%%%%%%%%%%%%%%%%%%%%%%%%%%%%%%%%

%%%%%%%%%%%%%%%%%%%%%%%%%%%%%%%%%%%%%%%%%%%%%%%%%%%%%%%%%%

\section{Introduction}

In any quantum theory of gravity, the low-energy effective action
is expected to include higher derivative corrections to the
Einstein-Hilbert action. In string theory, such corrections can be
computed in principle and in many cases may be found explicitly.
We will argue that these corrections play an important role in the physics
of certain black holes.  Many extremally charged black holes and branes have
null singularities that are not separated from asymptotic observers
by a regular horizon. For a broad class of examples, however,
the higher derivative terms completely alter the
causal structure of these objects, transforming these apparently singular
solutions into regular black holes with finite horizon area.  After this
`stringy cloaking', the near horizon geometry becomes $AdS_2
\times S^2$.

We will work in four dimensional, ungauged $N=2$ supergravity
\cite{deWvPr}.  In this case the BPS black hole solutions exhibit
fixed-point attractor behavior near the horizon, as discovered by
Ferrara,  Kallosh and Strominger \cite{FerKalStr,FK}.
Cardoso, de Wit and Mohaupt described how higher order derivative terms,
representing quantum effects, may be added to the action,
and found the generalized attractor equations
in the presence of these additional terms
\cite{CarDeWMoh}.
With Kappeli, they discussed the corresponding black hole solutions
\cite{CarDeWKMoh}.
We will follow the work of \cite{CarDeWMoh,CarDeWKMoh} closely
in what follows.
Wald has proposed a modification to the usual Bekenstein-Hawking
area law \cite{Haw} in the presence of higher curvature terms
\cite{Wald}, which was applied to these black holes by
\cite{CarDeWMoh}. This work focused primarily on black holes that
already have a regular horizon and large area in the leading
supergravity approximation, where the entropy receives small but
calculable corrections from higher derivative terms. Remarkably,
the leading order correction agrees with the counting of black
hole microstates performed by Maldacena, Strominger and Witten,
and by Vafa \cite{MSWV}. Recently, Ooguri, Strominger and Vafa
discussed an interesting relation between these black holes
and topological string theory \cite{OSV}.

The effects of the higher derivative terms are most dramatic for
black holes that classically have zero area  and hence zero
entropy. It was noticed recently by Dabholkar \cite{A} that for
certain two-charge black holes of this type occurring in Type-IIA
theory compactified on $K_3 \times T^2$, the
correction to the entropy coming from higher derivative 
prepotential-like
terms reproduces precisely the logarithm of the number of BPS
microstates
\cite{DH, DGHR}.

Motivated by these results, we examine the space-time geometry of
a general class of black holes in N=2 supergravity with higher
derivative corrections. In particular, we consider solutions of
Type-IIA string theory compactified on a general Calabi-Yau
3-fold, of which the two-charge black hole mentioned above is a
special case. We will focus on solutions which classically have
null singularities and zero horizon area. We will argue that in
these cases the higher curvature terms cause the black holes to
develop a regular horizon with non-zero area.\footnote{
This result is similar in spirit to the enhancon mechanism of Johnson, 
Peet and Polchinski \cite{jpp}, although the mechanism is quite different.}
In Einstein frame, the radius of the
horizon is large for large charges.
In addition to making the area non-zero, the higher curvature terms
significantly modify the Bekenstein-Hawking relation between entropy
and area. 
The net entropy $S$ is then related to the
corrected area $A$ by the relation $S=A/2$, which is distinctly
different from the standard relation $S=A/4$.

The paper is organized as follows. In $\S{2}$, we review the basic
formalism of the generalized attractor mechanism and describe black
hole solutions for
Calabi-Yau compactifications. More details can be found in
\cite{Mohaupt}. In $\S{3}$ we show that for many choices of electric
and magnetic charges the classical geometry has a null
singularity which is replaced by a smooth horizon once the
corrections are taken into account.  We discuss some
implications when the Calabi-Yau 3-fold is $K_3 \times
T^2$ in $\S{4}$,  and summarize our findings in $\S{5}$.

\section{Review}

%Before discussing quantum corrections to the area,
%we must summarize a few basic
%facts about these black hole solutions.
%For a more detailed review see \cite{Mohaupt}.

\subsection{Higher Curvature Corrections to $N=2$, $D=4$ Supergravity}

Our starting point is the $N=2$ superconformal theory of gravity
in four dimensions described by \cite{deWvPr}, which is
subsequently gauge fixed to give $N=2$ Poincar\'e supergravity.
The field content is a Weyl multiplet, which contains the graviton
as well as other physical and auxiliary fields, along with $n+1$
vector multiplets. One of these vector fields is the graviphoton,
which is not contained in the Weyl multiplet. One can define a
superfield ${\bf W}^{ij}_{ab}$, where $i,j=1,2$ are spinor indices
and $a,b$ are local Lorentz indices, which contains the field
strengths of the Weyl multiplet as well as an anti-selfdual
auxiliary field $T^{ab\, i j}$. The lowest component of the
associated chiral multiplet ${\bf W}^2 = ({\bf W}^{ab\,i j }
\varepsilon_{ij})^2$ is ${\hat A} = (T^{ab\, i j }
\varepsilon_{ij})^2$. The couplings between the vector multiplets
and  the Weyl multiplet are summarized by a holomorphic function
\be F(X^I,{\hat A}), \qquad I=0, 1,..., n \eq which is homogeneous
of degree two.  When $F$ is independent of $\hat A$, this is the
usual holomorphic prepotential, which depends on the scalar
components $X^I$ of the $n+1$ vector multiplets.

The supergravity field equations are covariant under symplectic
${\rm SP}(2n+2;{\rm \bf Z})$
transformations, which are a generalization of electric-magnetic duality.
These transformations act linearly on
the two $(2n+2)$-component vectors
\be
V =  \left( \begin{array}{c} X^I \\ F_I(X,{\hat A}) \\ \end{array} \right)
\;\;\;\;\; {\rm and} \;\;\;\;\;
 \left( \begin{array}{c} p^{ I} \\
q_{ I}  \\ \end{array} \right)\,.
\eq
Here $F_I={\partial F}/{\partial X^I}$, and
$p^I$, $q_I$ are the electric and magnetic charges of a given solution.
We can use these vectors to define the central charge
\be
Z =  e^{ {\cal K}/2} \, (p^I F_I (X,{\hat A}) - q_I X^I) \,,
\label{z}
\eq
where $e^{-\cal K} = i [\bar{X}^I F_I (X,{\hat A}) -
\bar{F}_I ({\bar X}, {\bar {\hat A}}) X^I ]$,
which is invariant under symplectic transformations.

When $F$ depends on $\hat A$ the action contains higher curvature
couplings between the Weyl and vector multiplets.
The bosonic Lagrangian contains the terms
\be
{8 \pi \cal L} = - \ft{1}{2} e^{-{\cal K}}R
+ \ft12( {i} F_{\hat{A}} \,\hat{C} +
\mbox{h.c.}) + \cdots \,,
\label{Paction}
\eq
where ${\hat C} = 64 \,C^{- \mu \nu \rho \sigma }
C^-_{ \mu \nu \rho \sigma}
+ 16 \,\varepsilon_{ij}\, T^{\mu \nu i j} f_{\mu}{}^{\rho\,} T_{\rho \nu k l}
\,\varepsilon^{kl} + \cdots \,$.  Here $C^-_{ \mu \nu \rho \sigma}$
denotes the anti-selfdual
part of the Weyl tensor $C_{ \mu \nu \rho \sigma}$, and
$f_{\mu}{}^{\nu} = \ft{1}{2} R_{\mu}{}^{\nu} - \ft{1}{12} R \,
\delta_{\mu}{}^{\nu}+ \cdots$.  To fix the local scale
invariance of the theory and obtain $N=2$ Poincar\'e supergravity,
we must set $e^{-{\cal K}}$ to one. The first term in the action is then the
properly normalized Einstein-Hilbert term.

\subsection{Black Hole Solutions}

Black hole solutions to this theory may be parameterized in terms
of two functions $f(r)$ and $g(r)$ as \beqa ds^2 = - e^{2g(r)}
dt^2 + e^{2f(r)} \Big[ dr^2 + r^2 ( \sin^2 \theta \, d\phi^2 + d
\theta^2 ) \Big] \;. \label{lineel} \eeqa At $r\rightarrow
\infty$, $e^{2g(r)}\rightarrow 1$ and $e^{2f(r)}\rightarrow 1$, so
the metric is asymptotically flat. As $r\rightarrow 0$, it was
shown by \cite{CarDeWMoh} that \beqa
 e^{ 2f(r)} = e^{-2g(r)}  =
\, { \vert Z\bar Z\vert_{r=0}\over   r^2}  \;
\label{f}\eeqa
where $Z$ is the central charge defined above.
When $Z|_{r=0}$ is non-vanishing
the solution becomes $AdS_2\times S^2$ at small $r$, and the black
hole has a finite horizon area
\be
A= 4 \pi \vert Z\bar Z\vert_{r=0} \ .
\label{area}
\eq
However, when
$\vert Z\bar Z\vert_{r=0}=0$ the solution has a null singularity: typically
$e^{ 2f(r)}=e^{-2g(r)}\sim 1/r$, and the horizon at $r=0$
is coincident with the null singularity.

Equation (\ref{area}) may also be derived by imposing
unbroken N=2 supersymmetry at the horizon.  Starting from
the gravitino variation
\be
\delta \psi_\mu^i = 2{\cal D}_\mu\epsilon^i
      -\ft18  T^{ab\,ij}\g_{ab}\,\g_\mu\epsilon_j 
\label{gravitino}
\eq
the gauge invariant condition 
$\delta {\cal D}_{[\mu}\psi_{\nu]}^i=0$ implies
a relation between the curvature of a near horizon $AdS_2\times
S^2$ and the value of the central charge $Z$ at $r=0$. This
procedure is explained in detail in \cite{enhan}.

%Rather than solving the full equations of motion,
To evaluate $Z|_{r=0}$, we may use the generalized attractor equations
\cite{CarDeWMoh}
\be
\bar{Z} \left( \begin{array}{c} X^I \\ F_I (X,{\hat A}) \\ \end{array} \right)
- Z \left( \begin{array}{c} \bar{X}^I \\ \bar{F}_I ({\bar X}, {\bar {\hat A}})
 \\ \end{array} \right)
= i
e^{-{\cal K}/2}
\left( \begin{array}{c} p^I \\ q_I \\ \end{array} \right) \,
\label{stab}
\eq
which follow from the constraint of unbroken N=2 supersymmetry at the horizon.
When there are no quantum corrections, i. e. when the prepotential $F$
is independent of ${\hat A}$, these are the usual attractor equations
\cite{FK}.
When $F$ depends on $\hat A$, consistency of the construction
requires that $\hat A =-  64\,e^{-{\cal K}}\,\bar Z^{-2}$.

The authors of \cite{CarDeWMoh} considered
regular black hole solutions which  classically have non-zero horizon area.
In the presence of higher curvature corrections,
the entropy is not given by the usual Bekenstein-Hawking formula.
Instead, following Wald's proposal \cite{Wald}, the authors of
\cite{CarDeWMoh} found
\be
S = \pi\,\Big[ \vert Z\vert^2 -256\,{\rm Im}\,[F_{\hat
A}(X^I ,\hat A)]\,\Big]\;, \quad\mbox{with}\quad {\hat A} =
-64\, \bar Z^{-2}\,{\rm e}^{-{\cal K}} \;\;,
\label{entropia}
\eq
where $X^I$ and  $Z$ are fixed by the
stabilization equations at $r=0$.
The first term is one quarter the horizon area (including higher derivative
corrections),
while the second term is the novel contribution to the entropy due to
Wald.

Following \cite{BCDWKLM}, we will introduce the rescaled variables
$Y^I = {\rm e}^{{\cal K}/2} {\bar Z} X^I$ and $\Upsilon = {\rm
  e}^{{\cal K}} {\bar Z}^2 {\hat A}$.
Using the homogeneity property of $F$, we find
\be
Z\bar Z =  p^I F_I( Y , \Upsilon) - q_I Y^I .
\eq
The stabilization equations (\ref{stab}) become
\be Y^I-\bar Y{}^I= i p^I \ ,
\eq
\be
F_I(Y,\Upsilon) -\bar F_I(\bar Y, \bar \Upsilon) = i
q_I \ .
\eq
The entropy formula (\ref{entropia}) takes the form
$S = \pi[ Z\bar Z - 256\, {\rm Im}\, F_{\Upsilon}
(Y,\Upsilon)]$ with $\Upsilon = -64$.

\subsection{Calabi-Yau Black Holes}

In this paper, we will
focus on black holes found by compactifying
type-IIA string theory on a large Calabi-Yau
threefold.
The prepotential has zero and linear order dependence on $\hat A$,
so that
\beqa
F(Y,\Upsilon)
= \frac{D_{ABC} Y^A Y^B Y^C}{Y^0} +
d_{A}\, \frac{Y^A}{Y^0} \; \Upsilon
\;\;.
\label{prepot}
\eeqa
Here $I = 0, \dots, n$, $A = 1, \dots, n$ and
\beqa
D_{ABC} = - \ft16 \,  C_{ABC} \;, \;
d_{A} = - \ft{1}{24} \, \ft{1}{64}\; c_{2A}
\label{prep2a}
\eeqa
depend on the Calabi-Yau
four-cycle intersection numbers $C_{ABC}$ and second Chern-class
coefficients $c_{2A}$
\cite{Bershadskyetal}.
The classical solution to the attractor equations,
where $\Upsilon\to 0$ in (\ref{prepot}), was found
by  Shmakova \cite{shma}.

The quantum corrected solution,
including the second term in (\ref{prepot}), is somewhat more complicated.
In this case the Lagrangian (\ref{Paction})
contains a higher curvature term proportional to
$c_{2A} \, z^A \,R^2$, where $z^A = Y^A/Y^0$.
Let us take $p^0 = 0$ and assume that there is a solution with $Y^0\ne 0$.
The solution to the generalized attractor equation was
found in this case to be \cite{CarDeWMoh}
\beqa
&& Y^0 = {\bar Y}^0 \;,\qquad  (Y^0)^2 =
\frac{D - 4 d_{A}\,
p^A \; \Upsilon}{4 {\hat q}_0} \, , \qquad  Y^A = \ft16 \, Y^0 \, D^{AB} q_B + \ft{1}{2} i\, p^A .
\label{ysol}
\eeqa
Here, following \cite{shma}, we denote
$ D\equiv  D_{ABC}\, p^A p^B p^C $,
$D_{AB} \equiv D_{ABC}\, p^C $, and define $D^{AB}$ by
$D_{AB} D^{BC} \equiv \delta_A^{\,C} $.
We have also introduced
${\hat q}_0=q_0 + \ft{1}{12} \, D^{AB}q_A q_B $, which we will take
to be negative. Note that this formula is valid only when
the matrix $D_{AB} = D_{ABC} p^C$ is invertible.
In the next section we will consider general configurations
where this may not be the case.
We must also take $p^A>0$ to keep $(Y^0)$ real.

Putting this together, we have the corrected formula for the horizon area
\beqa
{A}=  4 \pi|Z|^2_{r=0} = 16\pi Y^0 |{\hat q}_0| - 8\,\pi {d_A
\,  p^A\over Y^0} \,\Upsilon\,.
\eeqa
The corrected entropy formula, using $\Upsilon = -64$, is
\beqa
S = 4\pi\, Y^0 |{\hat q}_0|\;.
\eeqa
This agrees with the microscopic value computed in \cite{MSWV}.

\section{A Stringy Cloak for a Null Singularity}

In this section we study supersymmetric solutions with vanishing
classical horizon area, and conclude that quantum corrections due
to $R^2$-terms lead to  non-zero area.

Consider a Calabi-Yau black hole with
vanishing graviphoton magnetic charge $p^0=0$.
In the absence of higher curvature terms, i.e. taking $\Upsilon\to0$, the
area and entropy take the classical values
\beqa
S_{cl} = {A_{cl}\over 4}
    = 4\pi\, Y^0_{cl} |{\hat q}_0|\;
    = 2 \pi\, \sqrt{{\hat q_0} D_{ABC} p^A p^B p^C }.
\label{classa}
\eeqa
We have used here the attractor equations to fix
the classical value of $Y^0$ at $r=0$,
\be
(Y^0)^2_{cl} =
\frac{D }{4 {\hat q}_0} \, .
\label{yclass}
\eq

We will consider black holes with charges $p^A$ such that
\be
D\equiv  D_{ABC}p^A p^B p^C=0
\label{Dcond}
\eq
vanishes.  We see from (\ref{classa}) that the classical value for the
area vanishes.
Note that the classical value of $Y^0$ given by (\ref{yclass})
vanishes, so the attractor solutions presented above become
singular.

Fortunately, quantum corrections to the prepotential allow
one to consistently solve the attractor equations.
Making the ansatz
\beqa
Y^A =w^A + \frac{i}{2} p^A
\eeqa
where the $w^A$ are real, the attractor equations become
\beqa
D_{AB} w^B = \frac{1}{6} Y^0 q_A,~~~~~~~~
D_{AB} w^A w^B  = -\frac{1}{3} ( q_0 (Y^0)^2 + d_A p^A \Upsilon ).
\label{weq}
\eeqa
When $D_{AB}$ is invertible, the solution is given by (\ref{ysol}),
\be
Y^0 =
\sqrt{\frac{  c_{2A}\,
p^A }{24 |{\hat q}_0|}} \ .
\eq
The quantum corrected horizon area becomes
\beqa
{A}= 4 \pi  |Z|_{r=0}^2 =  8 \pi  \sqrt{\frac{  c_{2A}\,
p^A |{\hat q}_0|}{24 }} \ .
\label{corra}
\eeqa
This is non-zero provided $c_{2A} p^A$ and ${\hat q}_0$ are non-vanishing.
Note that $A$ depends on the topology of
the Calabi-Yau (via the Chern classes $c_{2A}$) as well as on the charges
of the black hole.  The area is large when ${\hat q}_0$, 
$c_{2A} p^A$ are large.  Formula (\ref{corra}) may change once
higher order $\alpha'$ and $g_s$ corrections are included.

If the matrix $D_{AB}$ is not invertible, then the solution is a
little more complicated.  In this case, as may be seen from
(\ref{weq}), one must impose additional conditions on the electric
charges $q_A$ to keep $Y^0$ non-zero. The simplest constraint is
$q_A=0$, in which case the above solution is correct provided we
set ${\hat q_0} = q_0$. More generally, the charge vector $q_A$
must be orthogonal to the kernel of the matrix $D_{AB}$.  In this
case we have ${\hat q_0} = q_0 + {1\over 12} {\hat D}^{AB}q_A q_B$,
where ${\hat D}^{AB}$ is the inverse of $D_{AB}$ in the subspace
orthogonal to it's kernel.

For all of the configurations described above, the corrected entropy is
\beqa
S =  4\pi\, Y^0 |{\hat q}_0|=4 \pi  \sqrt{\frac{  c_{2A}\,
p^A |{\hat q}_0|}{24 }} \ .
\eeqa
The quantum corrected
entropy-area relation for this class of black holes is
\beqa
S =  {A\over 2} \;.
\eeqa
We should emphasize that we have derived this relation only for this specific
class of solutions, where supersymmetry and
symplectic invariance strongly constrain the result.
In particular, both $S$ and $A$ are symplectic invariants, so it is not
surprising that they remain proportional even after
higher order corrections are included.
It would be interesting to
find the analogous formulae for other types of black holes.

\section{Two-charge Black Holes in $K_3\times T^2$ Compactification}

In this section we specialize to Type-IIA string theory
compactified on $K_3\times T^2$ considered in
\cite{A}. There are 23 two-cycles, provided by the torus $T^2$ and
the  22 two-cycles of $K_3$. The non-vanishing intersection
numbers are \be C_{1ab}= C_{ab} \ , \eq where $a=2,...,23$ runs
over the two-cycles of $K_3$ and $C_{ab}$ is the intersection
matrix on $K_3$. The second Chern class is $c_{2,1}=24$.

One virtue of this special example is that the power series
expansion of the holomorphic function $F(X^I,{\hat A})$ terminates
at first order in ${\hat A}$. Moreover, the full quantum corrected
expression for $F$ can be evaluated exactly (see e.g. \cite{A}).
This exact expression for $F$ includes the terms listed in
equation (\ref{prepot}), as well as corrections that are
suppressed when the volume of $K_3\times T^2$ is large. This
allows a more detailed comparison of the microscopic degeneracy
with the black hole partition function defined by \cite{OSV} to
all orders in the inverse charge in an asymptotic expansion.

We consider the configuration where $q_0=n$ and $p^1=w$ are
non-zero and the rest of the charges vanish. This corresponds to a
configuration of  $w$ D4-branes wrapping the $K_3$  with $n$
D0-branes sprinkled on the worldvolume. This state is dual to a
perturbative state with winding number $w$ and quantized momentum
$n$ in the heterotic theory compactified on $T^4 \times T^2$. 
The condition (\ref{Dcond}) is satisfied for this
configuration, so the black holes classically have zero area. The
quantum corrected area is \beqa
{A}%= 4 \pi  |Z|_{r=0}^2
    %=  8 \pi  \sqrt{\frac{  c_{2A}\,p^A |{\hat q}_0|}{24 }}
= 8\pi \sqrt {p^1|q_0|} \
\eeqa
and the corrected entropy is
\beqa
S %=  4 \pi  \sqrt{\frac{  c_{2A}\,p^A |{\hat q}_0|}{24 }}
=4 \pi \sqrt {p^1|q_0|}= {A\over 2}\; .
\eeqa

The corrected geometry of the black hole near the horizon is given
by $AdS_2 \times S^2$. In Einstein frame, 
the radius of the sphere goes as $N^{1\over
4}$, where $N \equiv |nw|$ can be taken to be large. The attractor
equations also determine the value of the heterotic dilaton $\phi$
at the horizon.  Recall that $e^{-2\phi}$ is given by the
imaginary part of $Y^1 / Y^0$, which gives $e^{-2\phi} \sim
\sqrt{N}$ from the solution of the attractor equations for our
charge configuration \cite{A}. The closed-string loop-counting
parameter $g_s^2 = e^{2\phi}$ thus varies in the black hole
geometry and reaches a constant value at the horizon that goes as
$1 \over \sqrt{N}$, and is thus vanishingly small for large N.

Given this value of the dilaton,  the radius of curvature at the
horizon is of order string scale in the string frame even
though it is large in Einstein frame. This is consistent with
the heuristic idea of a string-scale `stretched horizon'
\cite{Lenny} for the zero area extremal black holes \cite{Sen1} in
the heterotic description. Note that while the string metric
determines the sigma-model couplings, the Einstein metric is the
one that is invariant under duality.

In the Type-IIB description, this system is  dual to the D1-D5
system where the geometry is known to be desingularized  already
at the classical level without the inclusion of higher
curvature terms \cite{LM,LMM}. It was found that there 
are many such smooth geometries, which are dual to
the oscillating string solutions of \cite{CMP,Dabholkar}
that are specified by an arbitrary profile of left-moving
oscillations. The multiplicity of these smooth solutions then
accounts for the degeneracy of the corresponding BPS states. We
have found instead that the inclusion of higher curvature
terms generates a regular horizon, and the entropy associated with
this horizon accounts for the degeneracy of these states. It would
be interesting to understand in detail how these two dual pictures
are compatible with each other.

The entropy of these black holes no longer satisfies the
Bekenstein-Hawking  area formula but also includes the correction
due to higher derivative terms given by Wald's formula. For the
class of solutions considered here, we find that the entropy is
given by $S = A/2 \equiv A/4 + A/4$. Thus the correction to the
entropy is of the same order as the Bekenstein-Hawking entropy.
Note that $A$ here is the quantum corrected area and is quite
different from the classical area which vanishes.

This raises an important physical question. The Bekenstein-Hawking
formula for the entropy allows one to view the area theorems of
black hole dynamics in classical general relativity as special
cases of black hole thermodynamics. In particular, the fact that
the area of black holes always increases in  physical processes
can then be viewed as a special case of the second law of
thermodynamics that the entropy always increases in  physical
processes. For our black holes, since the entropy is quite
different from classical area, it is important to show that the
geometric quantity defined as the entropy by Wald's formula always
increases in a physical process. This is necessary to ensure that
the second law of thermodynamics is obeyed with the definition of
entropy  by Wald which appears in the first law of black hole
thermodynamics. Note that the entropy thus defined is in perfect
agreement with the counting of BPS microstates and the specific
corrections to the Einstein-Hilbert term used here are
well-motivated by string theory. Therefore,
slightly non-extremal black holes in this system could be a
good starting point for considering possible generalizations of
area theorems of classical general relativity within string
theory.

\section{Conclusions}

In conclusion, we have found a large  family of classically
singular, zero area solutions which are converted into regular
black holes by higher derivative corrections to the supergravity
action.  This class includes the two-charge solutions of string
theory on $K_3\times T^2$ studied recently in \cite{A}.

The entropy-area relation for this particular class of black holes
is $S = {A/2}$, which differs from the well known classical
relation  $S =   {A/4}$.  However, we expect that this
modified version the Bekenstein-Hawking formula applies only to
this special class of Calabi-Yau black holes.

It would be interesting to investigate other black hole solutions
with vanishing classical horizon area. 
We would also like to understand any relation between our stringy
cloaking device and the proposal of \cite{OSV},
as well as the effects of higher order $\alpha'$ and $g_s$ corrections
on our results.
We leave this for future work.

\vskip0.5cm

%\noindent
\leftline{\bf Acknowledgments} We would like to thank J. Harvey,
G. Horowitz, J. Hsu, S. Kachru, A. Linde, E. Martinec, S. Mathur,
J. McGreevy, M. Shmakova, E. Silverstein, S. Shenker, L. Susskind, 
R. Wald, and M. Zagermann for helpful discussions. This work was supported by
NSF grant 0244728, and by the Department of Energy under contract
number DE-AC02-76SF00515.

%\bibliography{ref}
%\bibliographystyle{utphys}

%\end{document}

%%%%%%%%%%%%%%%%%%%%%%
%%%%%%%%%%%%%%%%%%%%%%

\end{document}